\documentclass[%
 reprint,
 amsmath,amssymb,
pra,
]{revtex4-1}

\usepackage{graphicx}
\usepackage{dcolumn}
\usepackage{bm}
\usepackage{color}
\usepackage{epstopdf}
\usepackage{gensymb}
\usepackage[english]{babel}
\usepackage{float}
\usepackage{xr}
\usepackage{sidecap}
\usepackage{mathrsfs,amsmath}
\usepackage[normalem]{ulem}

\begin{document}

\preprint{APS/123-QED}

\title{Granular Graphene: direct observation of novel edge states on zigzag and armchair boundaries}

\author{L.-Y. Zheng}
\email{liyang.zheng.etu@univ-lemans.fr}
\author{F. Allein}
\author{V. Tournat}%
\author{V. Gusev}
\author{G. Theocharis}
 \email{georgiostheocharis@gmail.com}
\affiliation{%
LAUM, UMR-CNRS 6613, Le Mans Universit\'e, Av. O. Messiaen, 72085 Le Mans, France}%



\begin{abstract}

We propose a mechanical graphene analogue  which is made of stainless steel beads placed in a periodic magnetic field by a proper design. A stable, free of mechanical borders granular structure with well-predicted wave dynamics is experimentally constructed.
First we report the dispersion relation in conjunction with the evidence of the Dirac points. Theoretical analysis shows that, compared to genuine or other artificial graphene analogues, unconventional edge modes exist in the free zigzag and armchair boundaries together with novel bulk modes composed of in-plane extended translations but localized rotations at the edges. We observe the existence of edge modes in free zigzag boundary, and we reveal an experimental robust turning effect of edge waves from the zigzag to the armchair/zigzag boundary, even in the absence of a full band gap for bulk modes. 
Our work shows that granular graphene can serve as an excellent experimental platform to study novel Dirac, topological and nonlinear wave phenomena.

\end{abstract}
\maketitle

\section{Introduction}

Graphene, a single-layer of carbon atoms in honeycomb lattice, has recently emerged as an appealing system for conducting fundamental studies in condensed matter physics and in particular Dirac physics phenomena.   \cite{Neto, Sarma, Geim,Neto,Waka}. The great advances in experiments in graphene were awarded with the 2010 Nobel prize.
Despite the enormous progress, there are still great difficulties in designing/modifying graphene at will at the nanoscale. This leads many researchers to propose and study other artificial microscopic and even macroscopic graphene analogues for further fundamental studies.
These settings include the use of molecules \cite{review}, ultracold atoms \cite{ultracold}, photons \cite{RechtsmanNature,DuboisNC, ZhangPRL, AblowitzPRA} or phonons \cite{TorrentPRL,YuNM} in honeycomb lattice.

The study of edge wave in finite crystals has been a long  studied topic in condensed matter physics \cite{Shockley,Tamm}. 
In genuine graphene, zero-energy electronic states are predicted in nanoribbons with zigzag/beared boundaries \cite{Fujita,Nakada} and confirmed by means of scanning-tunneling microscopy \cite{Kobayashi,Niimi}.  However, armchair boundaries do not support electronic edge states unless defects appear on the edges \cite{Kobayashi} or the system is anisotropic \cite{Kohmoto}. 
The interest on edge states has been significantly renewed by recent advances in the study of topological physics. It has been shown that 
robust edge states/modes can appear in
topological insulators \cite{Hasan,Qi}, crystalline\cite{Fu} and higher order\cite{Benalcazar} topological insulators. 
In these systems, the existence of edge states is directly connected with the topological properties of bulk bands.
This is also the case for the electronic edge states of graphene, since they are related to the winding number of bulk eigenmodes \cite{Ryu,Mong,Delplace}. 

In addition to genuine graphene, an extensive body of works has been published over the last few years, studying  edge waves in different artificial graphene structures, particularly in photonics.
A number of edge modes has been observed experimentally, such as conventional flat bands as well as unconventional edge branches in zigzag and beared boundaries \cite{Kuhl,Plotnik,Milicevic}. Regarding armchair boundaries, the previous reports of armchair edge waves are either in anistropic microwave artificial graphene \cite{BellecNJP}, or in a photonic graphene-like structure of coupled micropillars \cite{MilićevićPRL}, where the existence of edge states is due to the coupling of $p_{x,y}$ photonic orbitals.
Considering the study of vibrational edge waves, the existence of edge modes has been predicted in genuine \cite{SavinPRB} and granular graphene \cite{ZhengEML}. Moreover, it has also been shown that  in mechanical graphene analogues both flat and dispersive unconventional edge modes can be found at zigzag edges under fixed boundary conditions \cite{Coriolis,Kariyado}. However, to the best of our knowledge, there is no experimental observation of edge waves in mechanical graphene till now.

In this work, we propose another type of artificial graphene, that is, granular graphene, which can be thought of as a mechanical analogue of graphene whose carbon atoms are replaced by macroscopic elastic beads and chemical bonds are substituted by contact interactions via various stiffnesses. Compared to genuine graphene or other mechanical graphene structures, granular graphene possesses extra physical features that make it very appealing from a fundamental point of view. One of those is the existence of multiple degrees of freedom (translations and rotations) \cite{MerkelPRL, HiraiwaPRL, PichardPRB, AlleinAPL}.
This, in combination with the honeycomb lattice geometry, leads to 
Dirac cones in the dispersion relation \cite{TournatNJP,ZhengUltrasonics} and topological helical edge waves \cite{ZhengPRB}. 
From an experimental standpoint, previous reports on two-dimensional (2D) granular crystals usually focus on closely-packed hexagonal or square lattices with mechanical constraints located on the borders \cite{GillesPRL, CostePRE2008,PrlChiara}.
Thus far however, no direct observation has been made on Dirac cones or edge wave propagation on granular graphene. The obstacles blocking further experimental investigations on granular systems include difficulties in constructing different structures and stability issues, particularly for looser packings like the honeycomb structure.
\begin{figure}[t]
\includegraphics[width=8.5cm]{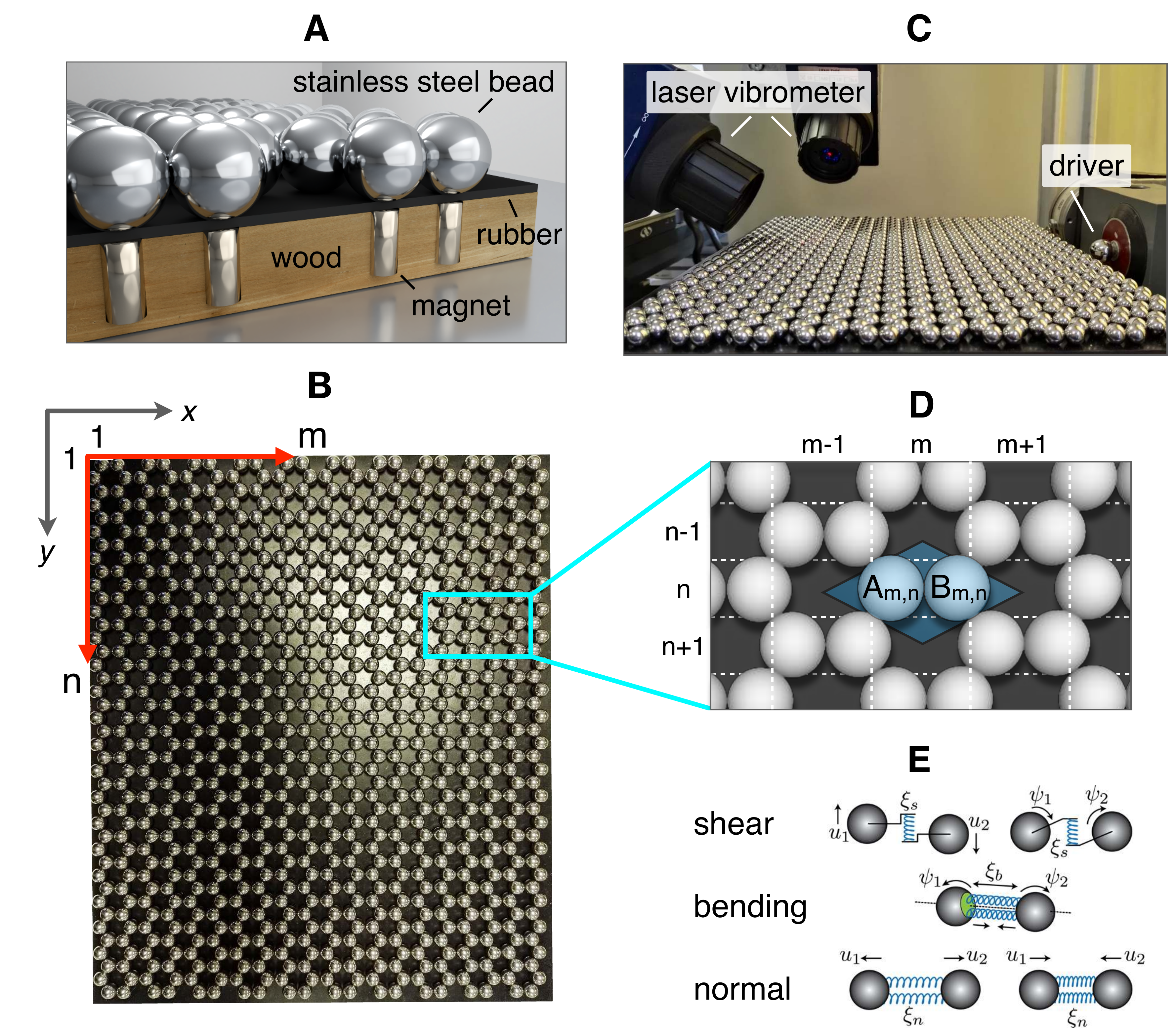}
\caption{\label{fig1} \textbf{Presentation of the Magneto-Granular Graphene (MGG).} (\textbf{A}) Cut-view schematics of the MGG where the magnetic field created by the permanent magnets induces attractive forces between the particles. (\textbf{B}) Top view of the realistic MGG, composed of 820 beads. (\textbf{C}) Experimental set-up for detecting wave propagation in the MGG. (\textbf{D}) Close-up schematic of the MGG. The blue box highlights a unit cell at position ($\mathrm{m},\mathrm{n}$) containing the two sublattice particles, labeled $A$ and $B$. (\textbf{E}) Considered interactions between beads in the MGG.}
\end{figure}

In this article, we  present how such structural difficulties have been overcome using external periodic magnetic fields. The proposed magneto-granular graphene (MGG) is structurally stable and acts as a nearly free-standing granular structure. We then experimentally show the dispersion curves, evidence of the Dirac point, and for the first time,  direct observations of the edge wave propagation in the MGG. 
We theoretically show that in the MGG, edge waves can exist not only on free zigzag, but also on free armchair boundaries. 
More importantly, for a range of frequencies we show the existence of novel bulk modes where translations are extended in the bulk of the structure while rotations are localized at the edges. In the same frequency range, when translations are also constrained at the edges, i.e., in partial band gaps of the bulk modes, edge modes can also appear on the free zigzag or both the free zigzag and armchair edges.
This leads to an interesting turning effect of edge waves from zigzag to armchair boundary in the frequency range where edge modes appear on both free zigzag and armchair edges.
Aside from the topological wave mechanism, where edge transport occurs in the full gap for bulk wave  and is protected by the bulk topology \cite{ZhuPRB}, the turning effect demonstrated here originates from the coexistence of wave solutions on the zigzag and armchair edges over a certain frequency range. The role of the topology in MGG, like in the recent works of higher order topological insulators, is a remaining intriguing question.

\section{Experimental set-up and modeling}
The MGG is depicted in Fig.~\ref{fig1}B, where $820$ stainless steel beads (diameter $d=7.95$~mm, density $\rho=7678$~kg/m$^3$, Young's modulus $E=190$~GPa and Poisson's ratio $\nu=0.3$) are precisely placed in a honeycomb lattice, 	in contact with one another.
This layout stems from a properly designed external magnetic field that is induced by permanent cylindrical NdFeB magnets (remanent magnetization $1.37$ T, diameter $6$ mm, and length $13$ mm) placed in a honeycomb configuration within the wood matrix, Fig.~\ref{fig1}A.
The external periodic magnetic field magnetizes the elastic beads, resulting in equivalent pre-compression forces between  beads and thus a mechanically stable structure. Between the elastic beads and the substrate, a thin layer of rubber (thickness $2$ mm) has been set
 to minimize the mechanical coupling of the granular graphene with the substrate, and to damp the transmission of elastic waves into the wood matrix. 
The experimental set-up is shown in Figs.~\ref{fig1}C. In-plane motion is excited by the driver connected to a piezoelectric transducer (\emph{Panametrics} V3052).
Each bead in the structure exhibits one out-of-plane rotation $\varphi$ around the $z-$axis and two in-plane translations $u$ and $v$ along the $x-$ and $y-$axes, respectively.
The $u$ and $v$ components of each bead can be monitored separately by two laser vibrometers, which are sensitive to changes in the optical path length along the beam direction.

\begin{figure*}[t]
\includegraphics[width=16.5cm]{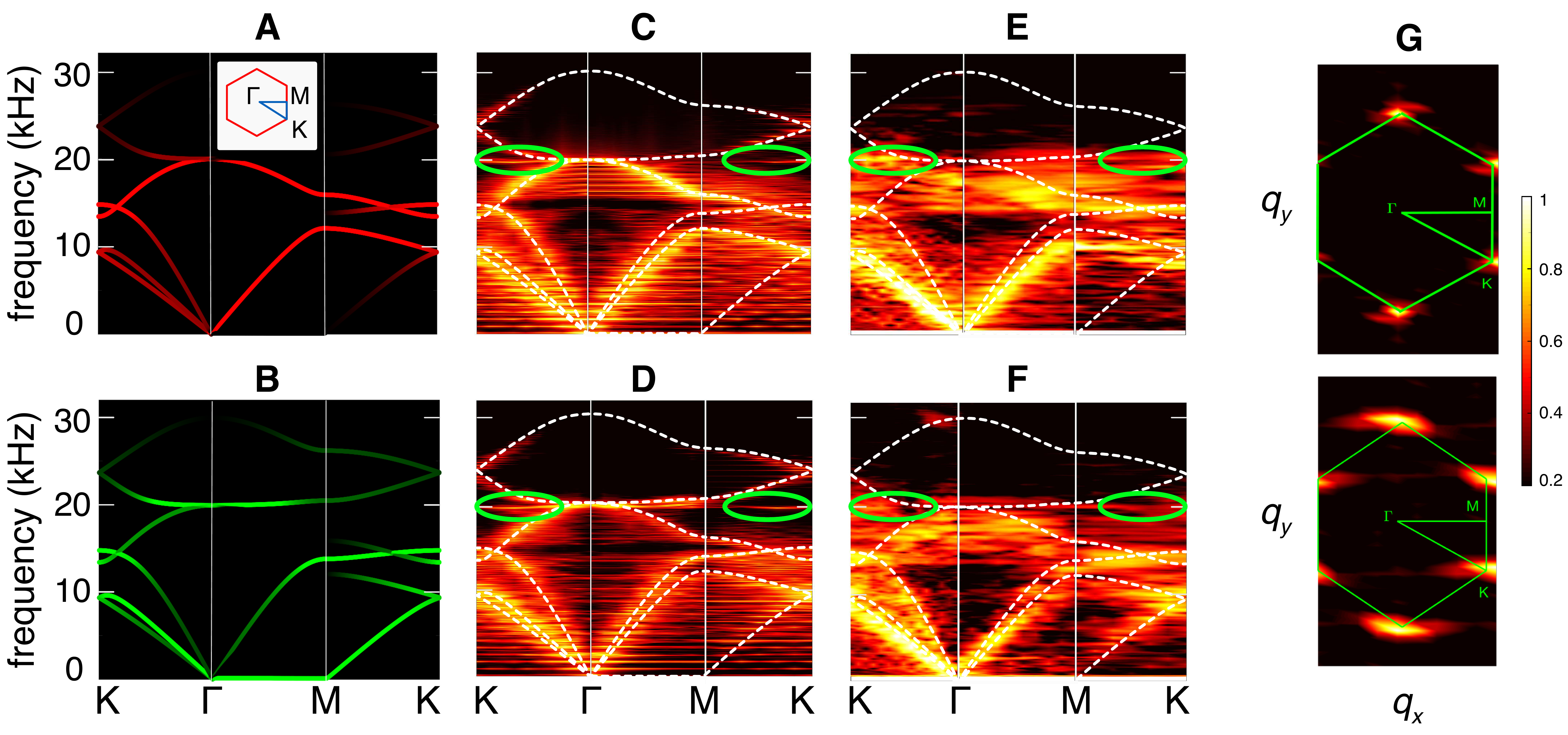}
\caption{\label{fig2} \textbf{Dispersion relations of the MGG and evidence of Dirac Point.} (\textbf{A}), (\textbf{B}) Theoretical (infinite MGG), (\textbf{C}), (\textbf{D}) numerical (finite MGG), (\textbf{E}), (\textbf{F}) experimental dispersion relations.  Panels (\textbf{A}), (\textbf{C}) and (\textbf{E}) show the dispersion curves of the $u$ component, while the $v$ component curves are presented in (\textbf{B}), (\textbf{D}) and (\textbf{F}).  (\textbf{G}) Iso-frequency contours of Dirac point: (Top) dirac point at $\omega_+ = 22.585$~kHz obtained from numerical simulation and (Bottom) the measured Dirac point at the frequency $\omega_+ = 21.731$~kHz. }
\end{figure*}

Regarding the mechanical contact interactions between adjacent beads, we consider normal, shear and bending interactions, as characterized by the contact rigidities $\xi_n$, $\xi_s$, and $\xi_b$  respectively, Fig.~\ref{fig1}E. Once pre-compression has been determined (around $\sim1.55$~N by means of measurement), $\xi_n$, $\xi_s$ and $\xi_b$ can be obtained from Hertzian contact mechanics \cite{Johnson, Mindlin}, see Methods.
For the three types of interactions between adjacent beads, i.e., Fig.~\ref{fig1}E, the elongations corresponding to the effective normal $n_\beta$, shear $s_\beta$, and bending $b_\beta$ contact  springs can be expressed as, 
\begin{subequations} \label{efs}
\begin{eqnarray} 
n_\beta&=&(u_{\beta}-u_\alpha)\boldsymbol e_x  \boldsymbol e_\beta+(v_{\beta}-v_\alpha)\boldsymbol e_y  \boldsymbol e_\beta, \\ 
s_\beta&=&(u_{\beta}-u_\alpha)\boldsymbol e_x  \boldsymbol l_\beta +(v_{\beta}-v_\alpha)\boldsymbol e_y \boldsymbol l_\beta -\dfrac{d}{2}(\varphi_\beta+\varphi_\alpha), \\
b_\beta&=&\dfrac{d}{2}(\varphi_\beta-\varphi_\alpha),
\end{eqnarray}
\end{subequations}
where $\alpha=A,B$ is the sublattice index. 
Considering the honeycomb structure, each sublattice bead is in contact with three other beads as denoted by a neighboring index $\beta=1,2,3$. We define $\boldsymbol e_\beta$ as unit vectors in the directions from the center of $\alpha$ bead to the center of its $\beta$-th neighbors. $\boldsymbol e_x$, $\boldsymbol e_y$ and $\boldsymbol e_z$ represent the unit vectors along $x-$, $y-$ and $z-$axes, respectively. $\boldsymbol l_\beta$ are unit vectors normal to $\boldsymbol e_\beta$ and $\boldsymbol e_z$ with the form $\boldsymbol l_\beta=\boldsymbol e_z\times\boldsymbol e_\beta$. 
As displayed in Fig.~\ref{fig1}D, we can label the sublattice $\alpha$ 
in a normalized coordinate $(\mathrm{m},\mathrm{n})$ (with the bead center positions serving as the coordinate) by $\alpha_{\mathrm{m},\mathrm{n}}$, 
where $\mathrm{m}$, $\mathrm{n}$ are both integers representing the normalized center positions of beads in the $x-$ and $y-$axes, respectively.
On site $(\mathrm{m}, \mathrm{n})$, the equations of motion can be expressed as follows,
\begin{subequations}\label{efm}
\begin{eqnarray}
M\ddot{u}_{\alpha,\mathrm{m},\mathrm{n}}&=&\sum_{\beta}(\xi_n n_\beta\boldsymbol e_x\boldsymbol e_\beta+\xi_s s_\beta\boldsymbol e_x\boldsymbol l_\beta), \\
M\ddot{v}_{\alpha,\mathrm{m},\mathrm{n}}&=&\sum_{\beta}(\xi_n n_\beta\boldsymbol e_y\boldsymbol e_\beta +\xi_s s_\beta \boldsymbol e_y \boldsymbol l_\beta), \\
I\ddot{\varphi}_{\alpha,\mathrm{m},\mathrm{n}}&=& \dfrac{d}{2}\sum_{\beta}(\xi_s s_\beta +\xi_b b_\beta).
\end{eqnarray}
\end{subequations}
Above, $M$ is the mass of a bead and $I$ is its moment of inertia. The dots on the top represent derivation over time. 
It can be seen from Eqs.~\eqref{efm} that bending interactions can not lead to the displacement of beads, i.e., Eqs.~(2a), (2b), while normal interactions do not contribute to the rotation of beads, i.e., Eq.~(2c). Based on the equations of motion in Eqs.~\eqref{efm}, wave dynamics in the MGG can be described by,
\begin{subequations}\label{eq1}
\begin{equation}
 \boldsymbol{\ddot{U}}_{\mathrm{m},\mathrm{n}}^A
=
S_0 \boldsymbol{U}_{\mathrm{m},\mathrm{n}}^A+
S_1 \boldsymbol{U}_{\mathrm{m},\mathrm{n}}^B+S_2
\boldsymbol{U}_{\mathrm{m-1},\mathrm{n+1}}^B+S_3
\boldsymbol{U}_{\mathrm{m-1},\mathrm{n-1}}^B,
\end{equation}
\begin{equation}
\boldsymbol{\ddot{U}}_{\mathrm{m},\mathrm{n}}^B
=
D_0 \boldsymbol{U}_{\mathrm{m},\mathrm{n}}^B+
D_1 \boldsymbol{U}_{\mathrm{m},\mathrm{n}}^A+D_2
\boldsymbol{U}_{\mathrm{m+1},\mathrm{n+1}}^A+D_3
\boldsymbol{U}_{\mathrm{m+1},\mathrm{n-1}}^A,
\end{equation}
\end{subequations} 
where $\boldsymbol U_{\mathrm{m},\mathrm{n}}^\alpha=[u_\alpha;v_\alpha;\Phi_\alpha]_{\mathrm{m},\mathrm{n}}$ 
with $\Phi=\varphi d/2$ are the motion of particle $\alpha$ in the normalized coordinates. $S_i$ and $D_i$ ($i=0,1,2,3$) are $3\times3$ matrices, see supplemental materials (SM). By applying the Bloch periodic boundary conditions in both $x-$ and $y-$axes, i.e., $\boldsymbol U_{\mathrm{m},\mathrm{n}}^\alpha=\boldsymbol U^\alpha e^{i \omega t -i q_x  \mathrm{m} - i q_y \mathrm{n} }$ with the normalized wave vectors $q_x=k_xd/2$, $q_y=\sqrt{3}k_yd/2$, 
Eqs.~\eqref{eq1} can be mapped into an eigenvalue problem which leads to the dispersion curves of an infinite MGG as shown in Figs.~\ref{fig2}A and ~\ref{fig2}B. 

Considering that the MGG in experiments is of a finite size $21\times41$, there are free zigzag edges at positions $(\mathrm{m},\mathrm{n})=(1,\mathrm{n})$, $(\mathrm{m},\mathrm{n})=(21,\mathrm{n})$ and free armchair edges at $(\mathrm{m},\mathrm{n})=(\mathrm{m},1)$, $(\mathrm{m},\mathrm{n})=(\mathrm{m},41)$.  At the mechanically free boundaries, which can be obtained by removing parts of the neighbors of edge beads, the beads are interacting with a smaller number of
neighboring beads than in the volume. Therefore, the boundary conditions are derived from
the cancellation of interactions between the removed beads and the edge beads, which leads to the following boundary conditions: 
\begin{subequations}\label{eq2}
\begin{equation}
M_0
\boldsymbol{U}_{1,\mathrm{n}}^B+D_1 \boldsymbol{U}_{1,\mathrm{n}}^A
= 0,
\end{equation}
\begin{equation}
M_1 \boldsymbol{U}_{21,\mathrm{n}}^A+S_1 \boldsymbol{U}_{21,\mathrm{n}}^B
=0,
\end{equation}
\end{subequations} 
for the zigzag edges, and,
\begin{subequations}\label{eq3}
\begin{equation}
M_2 \boldsymbol{U}_{\mathrm{m},1}^A+S_3
\boldsymbol{U}_{\mathrm{m}-1,0}^B
=0,
\end{equation}
\begin{equation}
M_3 \boldsymbol{U}_{\mathrm{m},1}^B+
D_3 \boldsymbol{U}_{\mathrm{m}+1,0}^A
=0,
\end{equation}
\begin{equation}
M_4 \boldsymbol{U}_{\mathrm{m},41}^A +
S_2 \boldsymbol{U}_{\mathrm{m}-1,42}^B
=0,
\end{equation}         
\begin{equation}
M_5 \boldsymbol{U}_{\mathrm{m},41} ^B+
D_2 \boldsymbol{U}_{\mathrm{m}+1,40}^A=0,
\end{equation}
\end{subequations}  
for the armchair edges with $M_j$ ($j=0,1,2,3,4,5$) $3\times3$ matrices, see SM. 
To account for dissipation, a phenomenological on-site damping term \cite{BoechlerNat} has also been introduced into the right-hand side of  Eqs.~\eqref{eq1},
 $-1/\tau\boldsymbol{\dot{U}}_{\mathrm{m},\mathrm{n}}^\alpha$, 
with $\tau$ characterizing the decay time of waves.
This coefficient has been chosen to fit the experimental results. 
More details about the dissipation implementation can be found in the Methods section.

\section{Dispersion curves and Dirac point}

To measure the MGG dispersion, in-plane motion has been activated using a frequency sweep excitation from $500$ Hz to $35$ kHz by the bead-driver located at position $(1, 22)$.
The $u$, $v$ components of particle $B$ in each unit cell are collected by the laser vibrometers. By scanning all particles $B$, the spatial frequency signals of translation can be obtained, which in turn yields the dispersion curves by applying a double Fourier transform.
Figures~\ref{fig2}A and \ref{fig2}B present the dispersion curves of an infinite granular graphene without dissipation.
The color scale level reflects the weights of $u$ (red curves) and $v$ (green curves) components in each mode.
The corresponding numerical dispersion curves, mimicking the experimental process, are displayed in Figs.~\ref{fig2}C-\ref{fig2}D, while the experimental ones are shown in Figs.~\ref{fig2}E-\ref{fig2}F for the $u$ and $v$ components respectively.
Figures~\ref{fig2}C-\ref{fig2}F indicate that up to $\sim 20$ kHz, the experimental dispersion curves are in good agreement with both the theoretical and numerical curves since the branches are translation-dominated.
As expected, the branches with frequencies above $\sim 20$ kHz are absent due to the fact that these modes are rotation-dominated, which are not easily detected by the laser vibrometers.

Interestingly, Figs.~\ref{fig2}E and \ref{fig2}F reveal the band crossing at the K point around frequency $10$ kHz. The observation of this crossing provides evidence of the Dirac cone in MGG, originating from the honeycomb lattice symmetry. As shown in Methods, there are theoretically two Dirac cones with the Dirac frequencies $\omega_\pm$ in granular graphene considering the in-plane motion. The band crossing around $10$ kHz corresponds to the Dirac point $\omega_-$ at the K point of the Brillouin zone (BZ).  Note that, 
another Dirac cone is also predicted around $\omega_+\sim 24$ kHz, Figs.~\ref{fig2}A-\ref{fig2}B. However, this Dirac point is not visible in Figs.~\ref{fig2}E and \ref{fig2}F due to the fact that the translation signals around $24$ kHz are weak and thus hidden in the color scale. 

In order to observe the $\omega_+$ Dirac point in the MGG, another set of experiment has been performed around the target frequency.
Experimentally we choose the source to be a frequency sweep excitation from $18$ kHz to $26$ kHz. 
However, there are still two main difficulties to be overcome: (1) Collection of the weak translational signals. Since the modes are dominated by rotation over this frequency region, the signal of translational components is consequently weak. Thus, this rotation-dominated modes are not easily detected by the laser vibrometers since they are only sensitive to changes due to displacements of beads. When dissipation is also taken into account, the weak translational motion becomes weaker due to the attenuation during propagation.
(2) The resolution of the Dirac point. Since the number of the eigenmodes is related to the size of the MGG, larger size of the structure results in a number of eigenmodes, which in turn lead to better resolution of dispersion around the Dirac point. However, as explained in (1), large size of the sample can lead to disadvantages for measurement due to the fact that the translational signals of the particles far away from the source can be too weak to be measured by the vibrometers. As a compromise, in this experiment for the $\omega_+$ Dirac point, we chose the size of sample to be $11\times41$. This keeps the resolution point along the $q_y$ unchanged, but decreases the length along $q_x$ to reduce the influence of  attenuation on the translational signal. 
By reconstructing a new sample of size $11\times41$ with stainless steel beads (diameter $8$~mm, density $7650$~kg/m$^3$, Young's modulus $210$~GPa and Poisson's ratio $0.3$), the pre-compression between beads in this sample is measured to be around $F_0=1$~N. Therefore, we calculate the Dirac frequency to be at $\omega_+\sim 22.585$ kHz. By scanning all the particles $B$ and doing the 2D real-reciprocal space Fourier transformation, iso-frequency contour of a given frequency can be obtained. In Fig.~\ref{fig2}G, we show the iso-frequency contours at the Dirac frequency $\omega_+$ obtained 
experimentally and numerically considering the same size of sample and dissipation. We observe that the mode at $\omega_+$, displayed by the high value in the iso-frequency contours, is only present around the K points which reveals an evidence of Dirac point in the MGG.
%

\begin{figure}[t!]
\includegraphics[width=8.0cm]{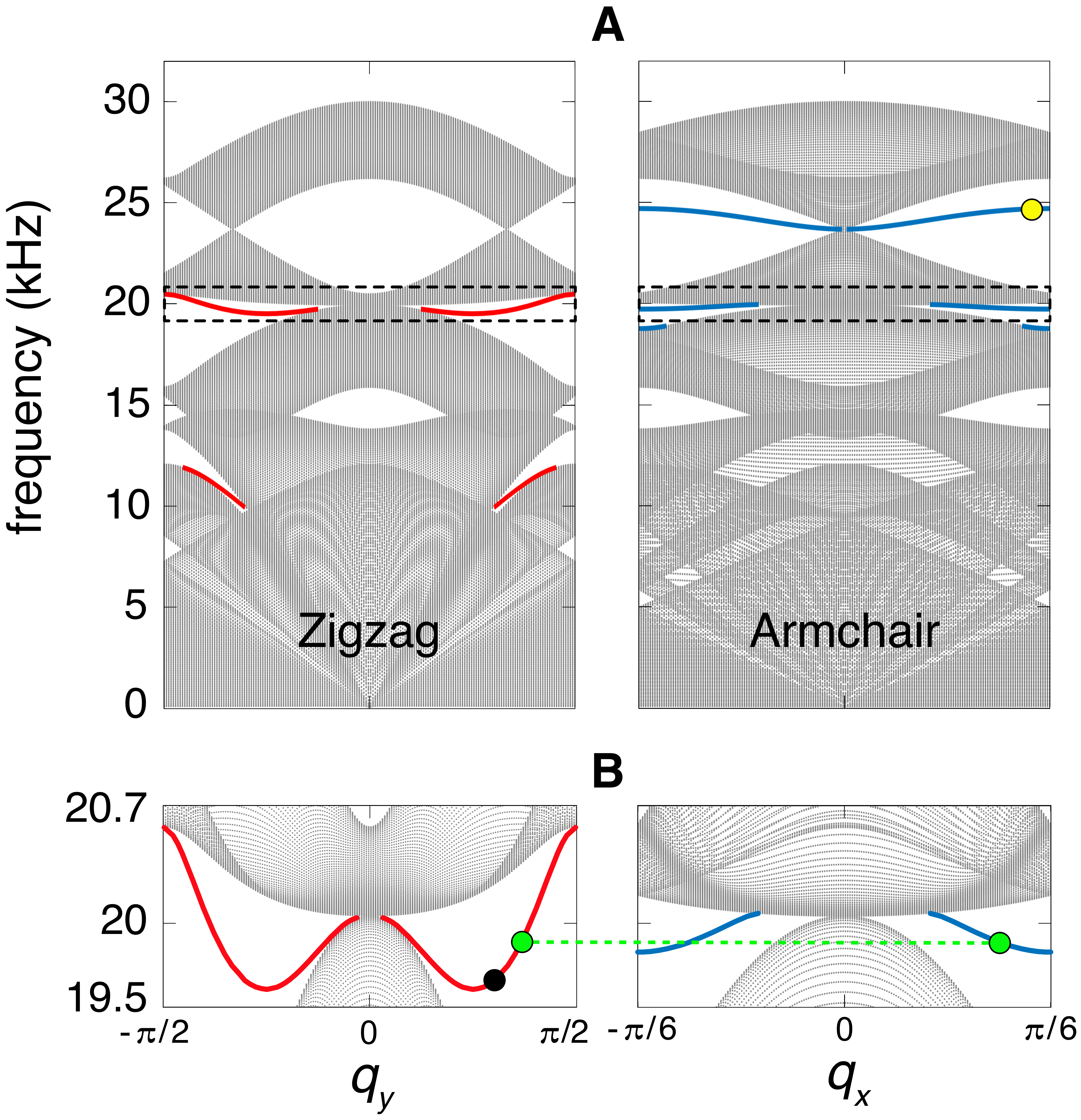}
\caption{\label{fig3} \textbf{Zigzag and armchair edge waves predictions.} (\textbf{A}) Edge wave dispersion curves for the zigzag and armchair edges. (\textbf{B}) Zoom around $20$~kHz. The gray regions represent the bulk modes, the red (blue) lines correspond to the edge wave branches. The modes marked in the edge branches at: $24.82$ kHz by the yellow dot in (\textbf{A}), $19.88$~kHz by the green dots in (\textbf{B}), and $19.62$~kHz by the black dot in (\textbf{B}) are displayed in Fig.~\ref{fig4}(\textbf{D-F}).}
\end{figure} 

\begin{figure*}[t!]
\includegraphics[width=16.5cm]{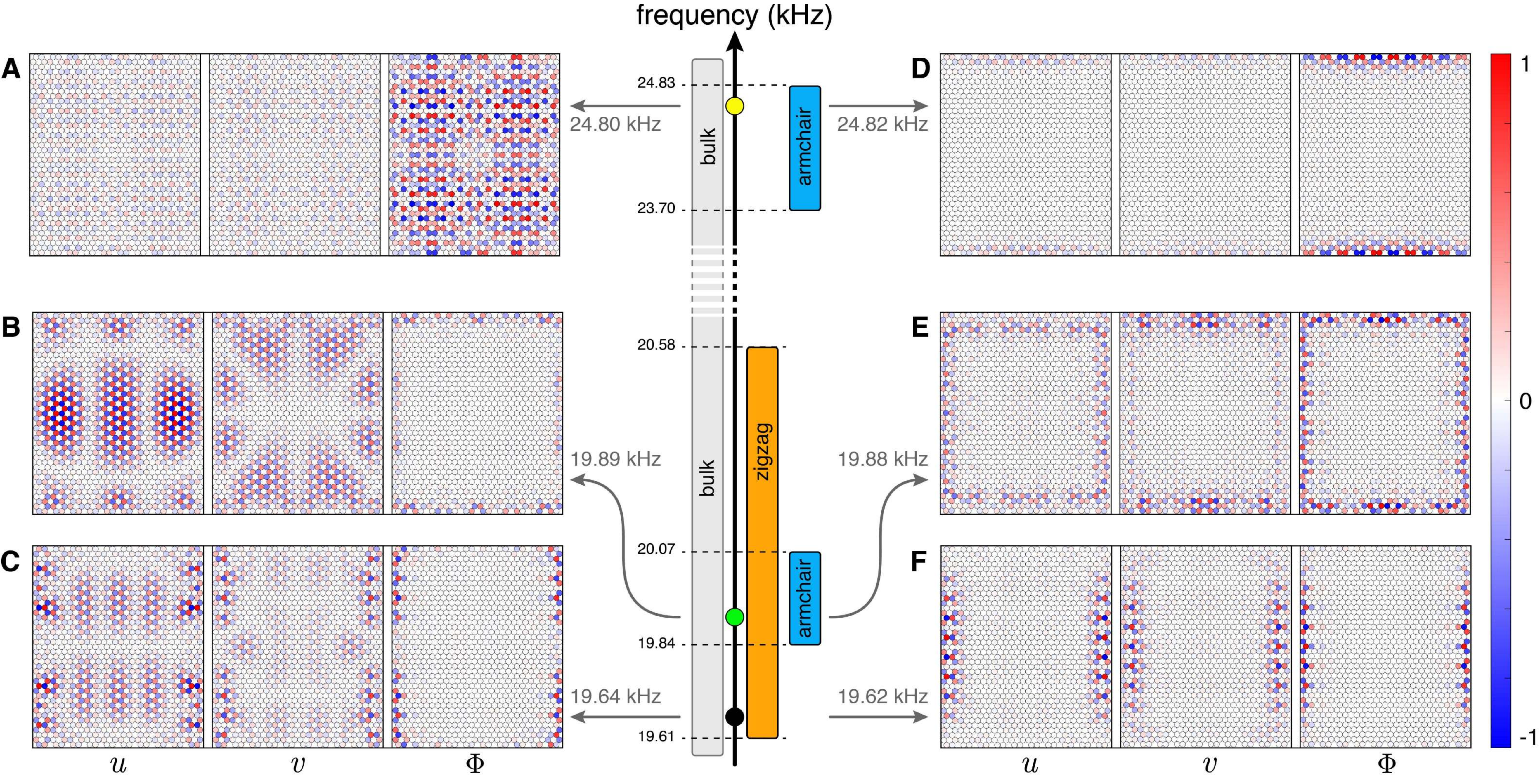}
\caption{\label{fig4} \textbf{Eigenmode analysis of the MGG.} Amplitude distributions of  the three components $u$, $v$ and $\Phi$ of edge eigenmodes (right panel \textbf{D-F}) for three different frequencies along with bulk mode (left panel \textbf{A-C}) having a frequency close to the edge modes.
The edge modes (\textbf{D-F}) correspond to the dots displayed in Fig.~\ref{fig3}.}
\end{figure*} 

\section{Edge waves}
Another interesting feature that appears in the dispersion of the finite-sized MGG is the existence of branches in the $\Gamma$K and MK directions around $20$ kHz (green ellipses in Fig.~\ref{fig2}C-\ref{fig2}F. These branches correspond to edge waves and in this section, we will study these first theoretically and then experimentally.

By considering free boundaries in Eqs.~\eqref{eq2} and~\eqref{eq3}, the edge wave dispersion for the zigzag and armchair edges is calculated, see Fig.~\ref{fig3}A. In the calculations of the edge dispersion, we assume that the free zigzag (armchair) edges are located at $\mathrm{m}=1$ and $\mathrm{m}=21$ ($\mathrm{n}=1$ and $\mathrm{n}=41$), while along the $y-$ ($x-$) axis are infinite. Therefore, based on Eqs.~\eqref{eq1} and the boundary conditions in Eqs.~\eqref{eq2} (Eqs.~\eqref{eq3}) along with the Bloch periodic conditions in the $y-$ ($x-$) axis, the edge wave dispersions in Fig.~\ref{fig3}A are obtained.
The gray regions correspond to bulk, while the red (blue) curves to edge wave solutions.
In total, $2$ edge branches for the zigzag and $3$ for the armchair are present.
This increased number of edge states, especially the existence of edge states at the armchair edge, is not encountered in the genuine graphene. 
Similar edge modes have been only reported in photonic lattices with orbital bands \cite{MilićevićPRL},  or mechanical honeycomb lattices with purely in-plane translations \cite{Coriolis,Kariyado}, namely 2D mass-spring honeycomb systems with two translational degrees of freedom per sublattice. However, in granular graphene, there are three degrees of freedom per sublattice (two translations $u$, $v$, one rotation $\phi$). As we will see, this significantly enriches the edge physics of this system.

Considering the edge wave in the rotation-dominated region (above $\sim 22$ kHz), as shown in Fig.~\ref{fig3}A, it can be seen that 
 no edge modes are found in the zigzag boundary. 
 This is different from the conventional graphene where a flat band of zero edge modes exists in the zigzag edges~\cite{Kohmoto, Ryu}. In mechanical graphene, similar flat band edge states can be found under
fixed boundaries conditions. As it was commented in Ref.~[\cite{Kariyado}], free boundary conditions, like the ones used in this work, break the chiral symmetry on the free zigzag edges, leading to the absence of a flat band of edge modes. However, a novel armchair edge branch appears, whose topological origin can be further discussed in future investigation.
In Fig.~\ref{fig3}B, we present a close-up of the edge wave dispersion around $20$ kHz for the zigzag and armchair cases, respectively. Interestingly, there is an overlapping region from $\sim 19.84$ kHz to $\sim 20.07$ kHz, where edge modes can be found on both the zigzag and armchair edges, while due to the absence of a full bulk gap, bulk modes also exist. 

To shed more light on the edge physics of the MGG, we carry on eigenmode analysis of the structure.
Since the MGG is of a finite size $21\times41$, a dynamical equation describing wave behavior of the MGG can be derived from Eqs.~\eqref{eq1}$-$\eqref{eq3} by taking into account all the coordinate indices $\mathrm{m}$ and $\mathrm{n}$. Consequently, the eigenvalue problem from the dynamical equation of the MGG can be thoroughly solved.
In Fig.~\ref{fig4}, we choose to show several eigenmodes around the edge wave frequencies. 
Starting from the rotation-dominated region, we show two eigenmodes with the eigen-frequencies $24.80$ kHz in Fig.~\ref{fig4}A, and $24.82$ kHz in Fig.~\ref{fig4}D. The color scales of the three components suggest that in this frequency region, modes are dominated by rotation. 
This is consistent with the measured dispersion curves in Figs.~\ref{fig2}E and ~\ref{fig2}F, where modes are not detected above $\sim 22$ kHz since rotational signals can not be recorded by the laser vibrometers.
It can be also seen that the eigenmode in Fig.~\ref{fig4}A shows an extended property since all parts of the structure are involved in the motion. However, the eigenmode in Fig.~\ref{fig4}D exhibits a property of localization as the motion mostly is confined only on the free armchair boundaries.
Note that, due to the boundaries, the eigenmodes of a finite-size graphene can be viewed as the contributions of those bulk and edge modes of the infinite MGG. Vibrational modes strongly localized at the edges of the structure can be called as edge modes of the MGG.
Therefore, the extended mode at $24.80$ kHz could be viewed as a mode dominated by the contribution from the bulk modes of infinite MGG of the grey region of Fig.~\ref{fig3}, while
the eigenmode in Fig.~\ref{fig4}D could be dominated by the mode marked by the yellow dot in Fig.~\ref{fig3}A in the armchair edge mode branch. 
Regarding the overlapping region, two eigenmodes close to each other at $19.89$ kHz and $19.88$ kHz are presented in Fig.~\ref{fig4}B and \ref{fig4}E, respectively. It shows that the eigenmode in Fig.~\ref{fig4}B also exhibits an extended property with translations involving most part of the structure, while the eigenmode in Fig.~\ref{fig4}E has a localized mode property similar to the one in Fig.~\ref{fig4}D but with motion constrained on both the zigzag and armchair boundaries. Thus, this eigenmode at $19.88$ kHz has a dominant contribution from the edge modes of infinite granular graphene in the zigzag and armchair edge branches labelled by green dots in Fig.~\ref{fig3}B.
The structure of this eigenmode confirms the prediction of the existence of edge branches in both zigzag and armchair ribbon (blue, red curves in Fig.~\ref{fig3}B). 
Finally, in the region that lies below the overlapping region ($\sim 19.61$ kHz to $\sim 19.84$ kHz), Fig.~\ref{fig4}C and~\ref{fig4}F, the behavior of eigemodes is quite similar to those in Fig.~\ref{fig4}B and~\ref{fig4}E but with the absence of motion in the armchair boundaries.  
The eigenmode at $19.62$ kHz, Fig.~\ref{fig4}F, is confined only on the zigzag edges, indicating the dominant contribution from the edge mode in the zigzag branch of the infinite graphene marked by the black dot in Fig.~\ref{fig3}B.

Another intriguing property observed in Fig.~\ref{fig4} is that the extended modes in the region from $\sim 19.61$ kHz to $\sim 20.58$ kHz manifest a very interesting behavior. As shown in Fig.~\ref{fig4}B,
the translational components $u,v$ of the eigenmodes are spread in the whole finite structure but the rotation is localized only in the boundaries. In Fig.~\ref{fig4}C, similar properties as the one in Fig.~\ref{fig4}B are observed but now the rotational component is only confined in the zigzag boundaries as the rotation of the edge mode in Fig.~\ref{fig4}F. This highlights a novel behavior of the dynamics in finite-size granular graphene, where one can find modes with extended translations in the structure while localized rotation on the boundaries.
To the best of our knowledge such a behavior has not been reported before in other graphene structures, and the rich wave physics originates here from the extra rotational degree of freedom. 
As a result, the rotation of the beads can have a very distinct behavior compared to the translation of the beads in the MGG. This can lead to interesting potential applications like rotation isolation devices in a more general mechanical system and advanced wave control.
 
\begin{figure*}[t]
\includegraphics[width=16.5cm]{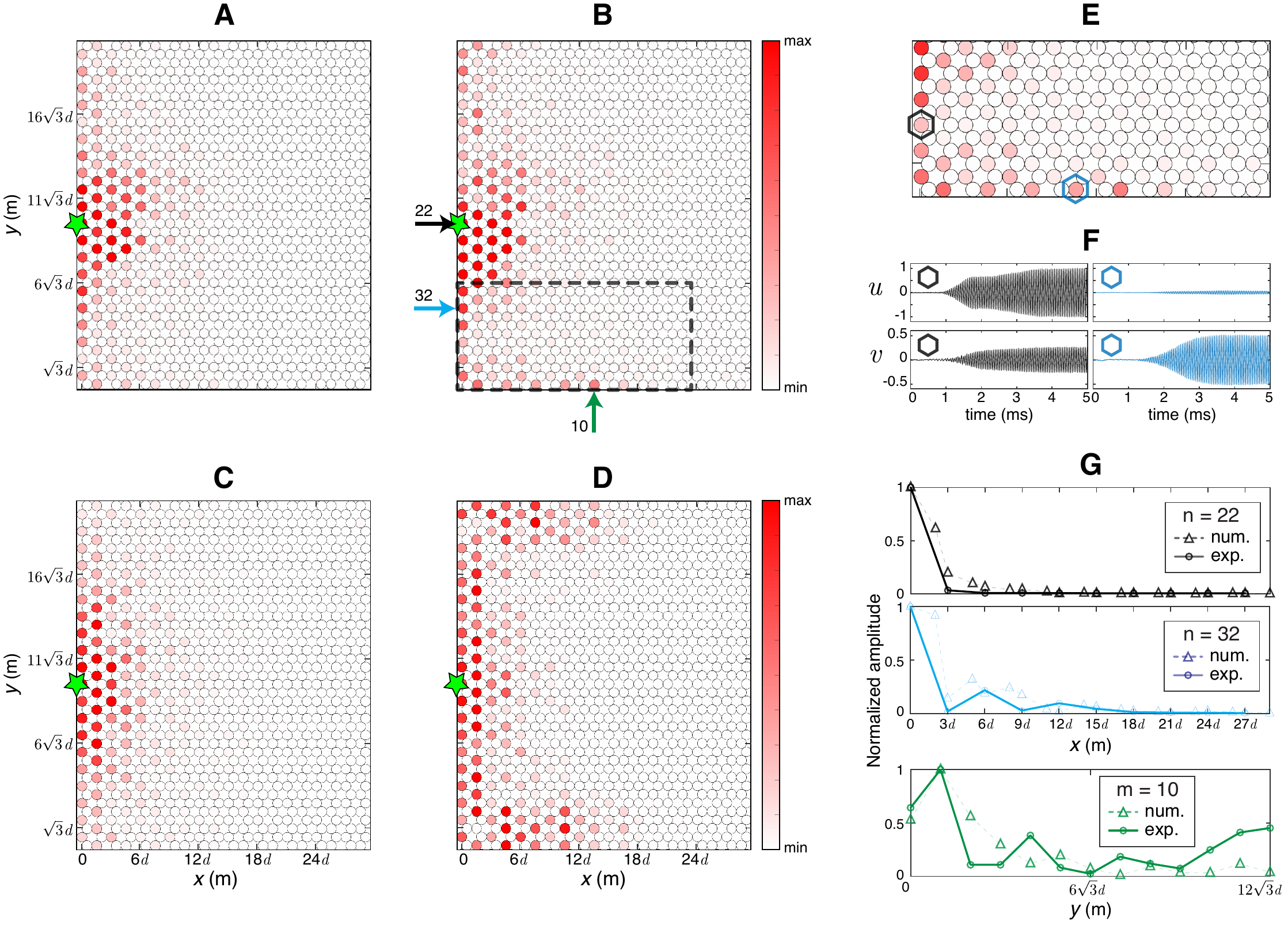}
\caption{\label{fig5} 
\textbf{Edge waves observation: turning effect of edge waves from the zigzag to the armchair boundary.} The spatiotemporal patterns of motion before the MGG reaches steady state are presented in (\textbf{A})  and (\textbf{C}), and after reaching steady state in (\textbf{B})  and (\textbf{D}). A harmonic wave at 20~kHz is excited from the source (green star).
The experimental results are shown in (\textbf{A}-\textbf{B}), while (\textbf{C}-\textbf{D})  are the simulations.
(\textbf{E})  Close-up of the corner marked by the dashed box in (\textbf{B}).
(\textbf{F})  The temporal signals of the two particles highlighted by black and blue hexagons in (\textbf{E}). (\textbf{G})  Translation distributions of beads along row 22 (black), row 32 (blue) and column 10 (green), depicted by the arrows in (\textbf{B}). }
\end{figure*}

The existence of edge waves in the MGG can be confirmed directly in experiment. To observe the edge wave propagation, the experimental set-up is the same as the one shown in Fig.~\ref{fig1}C, while a harmonic signal of duration $10$ ms with an initial linear ramping has been used as the source. 
All particles $B$ are still scanned to record the $u$, $v$ components. 
Figure~\ref{fig5} displays the measurements of total displacement amplitude ($\sqrt{u^2+v^2}$) for two separate times with a signal at $20$ kHz.
As shown in Fig.~\ref{fig5}A, when $t=1.7$ ms, the displacements are primarily localized on the zigzag edge while decaying into the bulk.
Despite the excitation of bulk waves at this frequency, decay of the bulk wave is expected due to both dissipation and 2D geometrical spreading
which provides a better observation of the elastic edge wave at $20$ kHz.
The numerical simulation of the experimental process is shown in Fig.~\ref{fig5}C, where the translational components of just the particles $B$ are shown. A good agreement between experiment and simulation is achieved.

\begin{figure*}[t]
\includegraphics[width=16.5cm]{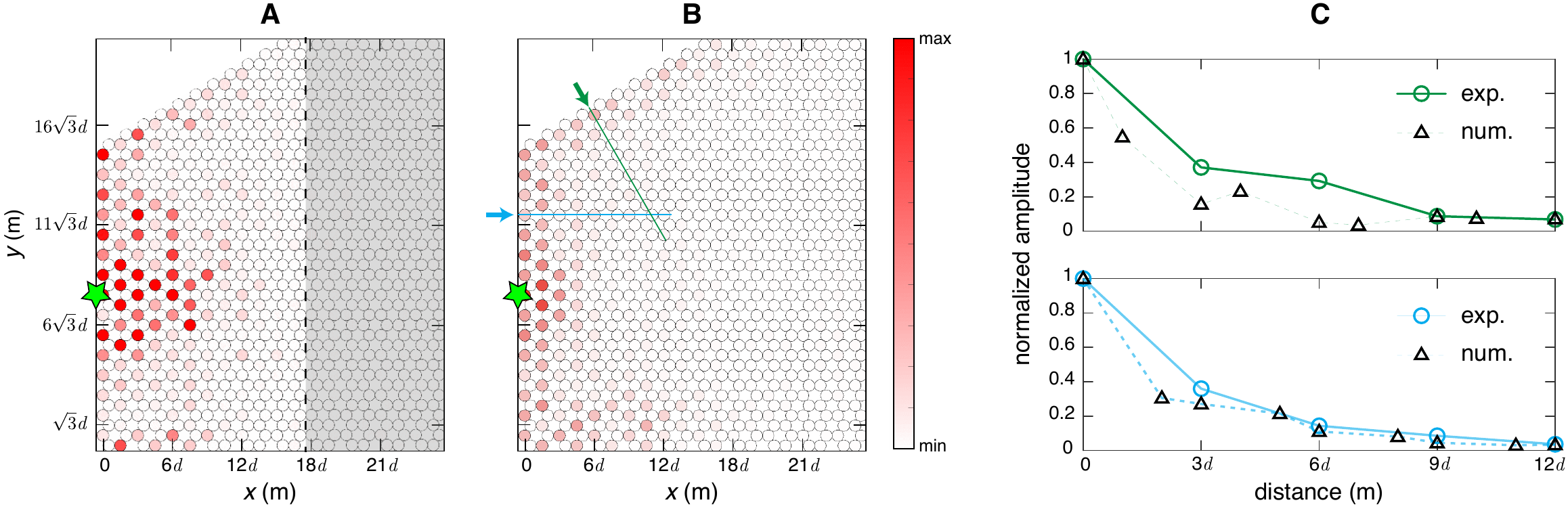}
\caption{\label{fig6} 
\textbf{Edge waves observation: turning effect of edge waves from the zigzag to the zigzag boundary.} The spatiotemporal patterns of motion for a modified MGG with zigzag to zigzag edges. A harmonic wave at 20~kHz is excited from the source (green star). The experimental result is shown in (\textbf{A}) where the gray zone corresponds to a region not experimentally scanned while (\textbf{B}) is the simulations. (\textbf{C}) Translation distributions of beads motion along the blue and green lines depicted by the arrows in (\textbf{B}).}
\end{figure*} 

\section{Novel turning effect of edge waves}
We now turn our attention to the frequency range at which edge modes coexist in both zigzag and armchair boundaries. For example, since $20$ kHz is located in this frequency range, one should expect 
that when the zigzag edge wave of $20$ kHz reaches the corner, this wave can be mode-converted into an armchair edge wave.
To observe this phenomenon, the spatial pattern at $t=3.7$ ms is depicted in Fig.~\ref{fig5}B and \ref{fig5}D. 
Indeed, wave motions are seen to be localized on both the zigzag and armchair boundaries. This can be further confirmed by the  close-up of the experimental spatial pattern of motion at the lower MGG corner as presented in Fig.~\ref{fig5}E. 
To demonstrate this turning effect more clearly, we have chosen two particles marked by black and blue hexagons in Fig.~\ref{fig5}E and we plot their time evolution in Fig.~\ref{fig5}F.
In addition, Fig.~\ref{fig5}G provides the spatial distribution of edge waves, which are obtained from experiments and simulations by focusing on rows $\mathrm{n}=22$, $\mathrm{n}=32$ and column $\mathrm{m}=10$, as labeled in Fig.~\ref{fig5}B by arrows.
In rows $\mathrm{n}=22$ and $\mathrm{n}=32$ the motion distribution of the mode shows a similar profile (the amplitude is normalized to the first bead on the left zigzag edge), confirming the edge mode property as the motion propagates along the edge while decays very fast into the bulk (a distance of around $x=9d$).
For $\mathrm{m}=10$, i.e. bottom panel of Fig.~\ref{fig5}G, the translational signal also reveals a localized profile close to the armchair boundary 
confirming that  the movement of beads in the armchair edge is due to the turning effect but not from the bulk modes.
Note that, as indicated in Figs.~\ref{fig5}B and \ref{fig5}D, due to dissipation and slow propagation velocity, the edge wave at $20$ kHz on the armchair edge is damped before propagating a long distance, e.g. $15d$. Further investigation of zigzag and armchair edge wave dynamics both with and without losses can be found in the SM. 
Note also that the spatial pattern in Fig.~\ref{fig5}B shows a small asymmetry of wave propagation in the upward and downward directions in the experiments.
This is most likely due to uncertainties in the pre-compression forces and asymmetric excitation of motion due to small misalignment between the driving bead and the set-up.

We now investigate the edge wave propagation along a corner of an angle of 120$\degree$, which connects a zigzag to another zigzag boundary.
Considering that the experimental MGG is still of a finite size $21\times41$, there is a free zigzag boundary at $(\mathrm{m},\mathrm{n})=(1,\mathrm{n})$. In order to build the second zigzag boundary, the particles above the line connecting the position $(\mathrm{m},\mathrm{n}) = (1,11)$ and $(\mathrm{m},\mathrm{n}) = (12,1)$ are removed, see Fig.~\ref{fig6}.

The experimental set-up for zigzag to zigzag edge wave measurement is the same as that shown in Fig.~\ref{fig1}C, where a harmonic signal of duration $10$ ms with an initial linear ramping  and a frequency of $20$ kHz has been used as the source at position $(\mathrm{m},\mathrm{n}) = (1,26)$. All particles $B$ are still scanned to record the $u$, $v$ components. Since edge modes are found on the zigzag edges at around $20$~kHz, one expects that when the zigzag edge wave reaches the 120$\degree$ corner, this wave can turn to the other zigzag boundary. To observe this phenomenon, Fig.~\ref{fig6}A displays the measurements of total displacement amplitude ($\sqrt{u^2+v^2}$) at a given time after reaching the steady state. Numerically, the simulation of the experimental process is shown in Fig.~\ref{fig6}B, where the translational components of only the particles $B$ are shown.

Indeed, wave motions are seen to be localized on the new zigzag edge and a good agreement between experiment and simulation is achieved. We can observe that bulk waves are also present because there is no gap for bulk mode at this frequency. In addition, Figure~\ref{fig6}C provides greater detail for the spatial distribution of zigzag edge waves at a given point of time by focusing on rows $\mathrm{n}=18$, and cut line from $(\mathrm{m},\mathrm{n}) = (5,8)$ to $(\mathrm{m},\mathrm{n}) = (9, 20)$, as labeled in Fig.~\ref{fig6}(B) by arrows. As shown in Fig.~\ref{fig6}C, the two profiles have a similar form with the translational signals becoming very weak (the amplitude is normalized to the first bead of each zigzag edge) after a distance of around $x=9d$. This confirms that bead movement in the new zigzag boundary is due to the turning effect from the zigzag to zigzag edges but not from the bulk modes.

\section{conclusions}

In this work, we propose a new artificial graphene called magneto granular graphene. This structure is composed of stainless steel beads in contact and placed in a properly designed magnetic field. The latter magnetizes the beads resulting in equivalent pre-compression forces between beads, and thus a mechanically stable structure, free of mechanical borders. The MGG proposed in this work can be used as a perfect experimental benchmark for fundamental study of Dirac and edge wave physics in mechanical systems. 

Considering the wave behavior in the MGG, first we obtain the dispersion relation and the Dirac points. Then, we turn our attention to the edge physics of the structure. We show that the MGG supports unconventional edge waves that can exist also in armchair free boundaries, in contrast with genuine or other artificial graphene. In addition we show that for a range of frequencies, the structure supports edge vibrations both on the zigzag and armchair boundaries. Interesting enough, in this region the bulk modes are extended in their translation motions but localized at the edges regarding their rotation motion. Such a unique behavior has not been reported before to the best of our knowledge. Novel applications such as rotational isolators could be then designed using MGG or other flexible mechanical metamaterials with rotational elements \cite{Qian, Babaee, Deng}.

Moreover, we also demonstrated that the coexistence of edge wave solutions in both zigzag and armchair boundaries lead to a turning effect from zigzag to armchair/zigzag free boundaries.
This does not require a full bulk gap, which normally is necessary in the scenario of pseudospin topologically-protected wave propagation, like the case of helical edge waves. The role of the topology in MGG, like in the recent works of higher order topological insulators, is a remaining intriguing question and might lead to potential study of novel topological phase in mechanical systems.
Finally, taking advantage of the intrinsic nonlinearities of the granular crystals, the MGG proposed herein offers a perfect platform to explore a wide array of novel nonlinear bulk, edge waves in mechanical graphene, similarly to solitons \cite{Nesterenko, Chong}, nonlinear waves \cite{CostePRE, JobPRL, LeonardExpMech, CabaretPRL} and breathers \cite{BoechlerPRL, TheocharisPRE} in simpler crystal structures.

\acknowledgments
This work has been funded by RFI Le Mans Acoustique in the framework of the APAMAS and Sine City LMac projects and by the project CS.MICRO funded under the program Etoiles Montantes of the Region Pays de la Loire and partly funded by the Acoustic Hub project.
\appendix
\section{Contact description in granular graphene}
Considering the in-plane motion in the magneto-granular graphene (MGG), there are normal, shear and bending interactions characterized by contact rigidities $\xi_n$, $\xi_s$ and $\xi_b$, respectively, as represented in Fig.~\ref{fig1}E. For stiffness of macroscopic elastic spheres in the MGG, the contact mechanism between the beads can be modeled by the Hertzian contact~\cite{Johnson, Mindlin}. This leads to the expressions of the rigidities:
\begin{subequations}\label{eq0}
\begin{eqnarray}
\xi_n &=& \left(\dfrac{3R}{4} F_0 \right)^{1/3}E^{2/3}(1-\nu^2)^{-2/3}, \\
\xi_s &=& \left(6 F_0 R \right)^{1/3}E^{2/3}\dfrac{(1-\nu^2)^{1/3}}{(2-\nu)(1+\nu)},
\end{eqnarray}
where $R$ is the radius of the bead, $E$ is the Young’s modulus, $\nu$ is Poisson’s ratio, and $F_0$ is the normal precompression between the beads. According to the previous study~\cite{ZhengUltrasonics}, the bending rigidity can be roughly estimated by,
\begin{eqnarray}
\xi_b &\sim &  \xi_n \left(\dfrac{\delta}{R}\right)^2, 
\end{eqnarray} 
where $\delta$ is the radius of the contact surface between two beads, which is given by,
\begin{eqnarray}
\delta &=& \left(\dfrac{3R}{4E}F_0\right)^{1/3}(1-\nu^2)^{1/3}.
\end{eqnarray} 
\end{subequations}
As long as the physical parameters of the beads are known and the precompression $F_0$ is measured, $\xi_n$, $\xi_s$ and $\xi_b$ can be obtained. In our experiment, the stainless steel beads have the parameters: Young’s modulus $E= 190$ GPa, Poisson’s ratio $\nu=0.3$, diameter $d=7.95$ mm, and density $\rho=7678$ kg/m$^3$. The procompression can be determined around $F_0 \sim 1.55$ N by means of measurement. This leads to the rigidities: $\xi_n\simeq 6.19\times10^{6}$ N/m,  $\xi_s\simeq 5.09\times10^{6}$ N/m, and $\xi_b\simeq 3.04\times10^{2}$ N/m.

\section{Calculation of Dirac points}
Let us consider the modes at the corner  (K point) of the BZ, by applying the periodic boundary condition, i.e., $\boldsymbol U_{\mathrm{m},\mathrm{n}}^\alpha=\boldsymbol U^\alpha e^{i \omega t -i q_x  \mathrm{m} - i q_y \mathrm{n} }$, Eqs.~\eqref{eq1} lead to two degenerate modes at the K point:\\
$\omega_{D_\pm}=\sqrt{\dfrac{g\pm \sqrt{g^2-h}}{4M}}$.\\
Above, $g=3[\xi_n+\xi_s+2\mathrm{P}(\xi_b+\xi_s)]$, and $h=72\mathrm{P}(\xi_n\xi_b+\xi_n\xi_s+\xi_b\xi_s)$ with $\mathrm{P}=MR^2/I$. 
These degenerate modes originate from the symmetry of honeycomb lattice, which are two Dirac points with frequencies $\omega_{D_\pm}$ at the K point. 


\section{Dissipation} 
In order to compare the experimental results with the numerical simulations, the attenuation of the wave during the propagation has to be considered. In our model, the attenuation is implemented by a phenomenological linear viscous on-site dissipation \cite{BoechlerNat} considering a time of decay, $\tau$, for elastic waves which can take different values as a function of displacement polarization. This leads to extra terms in the right hand side of Eqs (\ref{eq1}),
\begin{subequations}\label{eq_dissipation}
\begin{eqnarray}
 \boldsymbol{\ddot{U}}_{\mathrm{m},\mathrm{n}}^A
& = &
S_0 \boldsymbol{U}_{\mathrm{m},\mathrm{n}}^A+
S_1 \boldsymbol{U}_{\mathrm{m},\mathrm{n}}^B+S_2
\boldsymbol{U}_{\mathrm{m}-1,\mathrm{n}+1}^B  \nonumber \\ 
& & +S_3 \boldsymbol{U}_{\mathrm{m}-1,\mathrm{n}-1}^B
 - \frac{1}{\tau}  \boldsymbol{\dot{U}}_{\mathrm{m},\mathrm{n}}^A,
\end{eqnarray}
\begin{eqnarray}
\boldsymbol{\ddot{U}}_{\mathrm{m},\mathrm{n}}^B
& = &
D_0 \boldsymbol{U}_{\mathrm{m},\mathrm{n}}^B+
D_1 \boldsymbol{U}_{\mathrm{m},\mathrm{n}}^A+D_2
\boldsymbol{U}_{\mathrm{m}+1,\mathrm{n}+1}^A \nonumber \\ 
& & +  D_3
\boldsymbol{U}_{\mathrm{m}+1,\mathrm{n}-1}^A  - \frac{1}{\tau}  \boldsymbol{\dot{U}}_{\mathrm{m},\mathrm{n}}^B.
\end{eqnarray}
\end{subequations} 
Therefore, wave dynamics of the MGG considering dissipation can be described by combining the boundary conditions in Eqs.~(4) and Eqs.~(5) with Eqs.~(\ref{eq_dissipation}). We can notice that Eqs.~(\ref{eq_dissipation}) are second order ordinary differential equations of time. As a consequence, we numerically obtain the time evolution of elastic wave propagation in the structure by solving Eqs.~(\ref{eq_dissipation}) using Runge Kutta fourth order method. More details about the time evolution of elastic wave propagation in two-dimensional granular crystals can be found in Ref.\cite{ZhengPRB}. By fitting the experimental results with the numerical ones, we can estimate that $\tau$ is about 1 ms for both polarizations of displacement in our experiment.
%



\begin{thebibliography}{99}
\bibitem{Neto}
A. H. C. Neto, F. Guinea, N. M. R. Peres, K. S.Novoselov, and A. K. Geim, The electronic properties of graphene. Rev. Mod. Phys. \textbf{81}, 109 (2009).

\bibitem{Sarma}
S. D. Sarma, S. Adam, E. H. Hwang, and E. Rossi, Electronic transport in two-dimensional graphene. Rev. Mod. Phys. \textbf{83}, 407 (2011).

\bibitem{Geim}
A. K. Geim, and K. S. Novoselov, The rise of graphene. Nature Mater. \textbf{6}, 183–191 (2007).

\bibitem{Waka}
K. Wakabayashi, Ken-ichi Sasaki, T. Nakanishi and T. Enoki, Electronic states of graphene nanoribbons and analytical solutions. Sci. Technol. Adv. Mater. \textbf{11} 054504 (2010).

\bibitem{review} 
M. Polini, F. Guinea, M. Lewenstein, H. C. Manoharan, and V. Pellegrini, Artificial honeycomb lattices for electrons, atoms and photons. Nature Nanotechnology, {\bf 8}, 625 (2013).

\bibitem{ultracold}
L. Tarruell, D. Greif, T. Uehlinger, G. Jotzu, and T. Esslinger, Creating, moving and merging Dirac points with a Fermi gas in a tunable honeycomb lattice. Nature {\bf 496}, 302 (2012).

\bibitem{RechtsmanNature}
M. C. Rechtsman, J. M. Zeuner, Y. Plotnik, Y. Lumer, D.
Podolsky, F. Dreisow, S. Nolte, M. Segev, and A. Szameit,
Photonic Floquet topological insulators. Nature (London) {\bf 496}, 196 (2013).

\bibitem{DuboisNC}
M. Dubois, C. Shi, X. Zhu, Y. Wang, and X. Zhang, Observation of acoustic Dirac-like cone and double zero refractive index. Nat. Commun. {\bf 8}, 14871, (2017).

\bibitem{ZhangPRL}
X. Zhang,  and Z. Liu, Extremal transmission and beating effect of acoustic waves in two-dimensional sonic crystals. Phys. Rev. Lett. {\bf 101}, 264303 (2008).

\bibitem{AblowitzPRA}
M. J. Ablowitz, S. D. Nixon, and Y. Zhu, Conical diffraction in honeycomb lattices. 
Phys. Rev. A {\bf 79}, 053830  (2009).


\bibitem{TorrentPRL}
D. Torrent, and J. Sanchez-Dehesa, Acoustic Analogue of Graphene: Observation of Dirac cones in acoustic surface waves. Phys. Rev. Lett. {\bf 108}, 174301 (2012).

\bibitem{YuNM}
S.-Y. Yu, X.-C. Sun, X. Ni, Q. Wang, X.-J. Yan, C. He, X.-P. Liu, L. Feng, M.-H. Lu, and Y.-F. Chen, Surface phononic graphene.
Nature Mater. {\bf 15}, 1243–1247 (2016).

\bibitem{Shockley}
W. Shockley, On the surface states associated with a periodic potential. Phys. Rev. 56, 317 (1939).

\bibitem{Tamm} I. Tamm, Uber eine mogliche art der elektronenbindung an kristalloberflachen. I. Z. Physik 76, 849 (1932). https://doi.org/10.1007/BF01341581.

\bibitem{Fujita} 
Fujita, K. Wakabayashi, K. Nakada, and K. Kusakabe, Peculiar localized state at zigzag graphite edge. J. Phys.
Soc. Jpn. 65, 1920 (1996).

\bibitem{Nakada} Nakada, M. Fujita, G. Dresselhaus, andM. S. Dresselhaus, Edge state in graphene ribbons: Nanometer size effect and edge shape dependence. Phys.
Rev. B 54, 17954 (1996).

\bibitem{Kobayashi}
Y. Kobayashi, K. Fukui, T. Enoki, K. Kusakabe, and Y. Kaburagi, Observation of zigzag and armchair edges of graphite using scanning tunneling microscopy and spectroscopy. Phys. Rev. B \textbf{71}, 193406 (2005).

\bibitem{Niimi} 
Niimi, T. Matsui, H. Kambara, K. Tagami, M. Tsukada, and
H. Fukuyama, Scanning tunneling microscopy and spectroscopy of the electronic local density of states of graphite surfaces near monoatomic step edges. Phys. Rev. B 73, 085421 (2006).

\bibitem{Kohmoto}
M. Kohmoto, and Y. Hasegawa, Zero modes and edge states of the honeycomb lattice. Phys. Rev. B \textbf{76}, 205402 (2007).

\bibitem{Hasan} 
M. Z. Hasan and C. L. Kane, Topological Insulators. Rev. Mod. Phys. 82, 3045
(2010).

\bibitem{Qi} 
X.-L. Qi and S.-C. Zhang, Topological insulators and superconductors. Rev. Mod. Phys. 83, 1057 (2011).

\bibitem{Fu} 
L. Fu, Topological Crystalline Insulators. Phys. Rev. Lett. 106, 106802 (2011).

\bibitem{Benalcazar} 
W. A. Benalcazar, B. A. Bernevig, and T. L. Hughes,
Quantized electric multipole insulators. Science 357, 61 (2017).

\bibitem{Ryu}
S. Ryu and Y. Hatsugai, Topological origin of zero-energy edge states in particle-hole symmetric systems. Phys. Rev. Lett. \textbf{89}, 077002 (2002).

\bibitem{Mong}
R. S. K. Mong and V. Shivamoggi, Edge states and the bulk-boundary correspondence in Dirac Hamiltonians. Phys. Rev. B 83, 125109 (2011).

\bibitem{Delplace}
P. Delplace, D. Ullmo, and G. Montambaux, Zak phase and the existence of edge states in graphene. Phys. Rev. B 84, 195452 (2011).

\bibitem{Kuhl}
U. Kuhl, S. Barkhofen, T. Tudorovskiy, H.-J. Stockmann, T. Hossain, L. de Forges de Parny, and F. Mortessagne, Dirac point and edge states in a microwave realization of tight-binding graphene-like structures. Phys. Rev. B. 82, 094308 (2010).

\bibitem{Plotnik}
Y. Plotnik, M.C. Rechtsman, D. Song, M. Heinrich, J.M. Zeuner, S. Nolte, Y. Lumer, N. Malkova, J. Xu, A. Szameit, Z. Chen, and M. Segev, Observation of unconventional edge states in 'photonic graphene'. Nature Mater. \textbf{13}, 57–62 (2014).

\bibitem{Milicevic}
M. Mili\'cevi\'c, T. Ozawa, P. Andreakou, I. Carusotto, T. Jacqmin, E. Galopin, A. Lemaître, L. Le Gratiet,
I. Sagnes, J. Bloch and A. Amo, Edge states in polariton honeycomb lattices. 2D Mater. 2 034012 (2015).

\bibitem{BellecNJP}
M. Bellec, U. Kuhl, G. Montambaux, and F. Mortessagne, The existence of topological edge states in honeycomb plasmonic lattices. New J. Phys. {\bf 16}, 113023 (2014).

\bibitem{MilićevićPRL}
M. Mili\'cevi\'c, T. Ozawa, G. Montambaux, I. Carusotto, E. Galopin, A. Lemaître,
L. Le Gratiet, I. Sagnes, J. Bloch, and A. Amo, Orbital edge states in a photonic honeycomb lattice. Phys. Rev. Lett. {\bf 118}, 107403 (2017).

\bibitem{SavinPRB} A. V. Savin, and Y. S. Kivshar, Vibrational Tamm states at the edges of graphene nanoribbons. Phys. Rev. B {\bf 81}, 165418 (2010).

\bibitem{ZhengEML} L.-Y. Zheng, V. Tournat, and V. Gusev, Zero-frequency and extremely slow elastic edge waves in mechanical granular graphene. Extreme Mechanics Letters {\bf 12}, 55–64 (2017).

\bibitem{Coriolis}
Y.-T. Wang, P.-G. Luan, and S. Zhang, Coriolis force induced topological order for classical mechanical vibrations. New J. Phys. {\bf 17}, 073031 (2015).

\bibitem{Kariyado} 
T. Kariyado and Y. Hatzugai,  Manipulation of Dirac cones in mechanical graphene. Sci. Reports  {\bf 5}, 18107 (2015).

\bibitem{MerkelPRL} A. Merkel, V. Tournat, V. Gusev, Experimental Evidence of Rotational Elastic Waves in Granular Phononic Crystals. Phys. Rev. Lett. {\bf 107}, 225502 (2011).

\bibitem{HiraiwaPRL} M. Hiraiwa, M. Abi Ghanem, S. P. Wallen, A. Khanolkar, A. A. Maznev, and N. Boechler, Complex contact-based dynamics of microsphere monolayers revealed by resonant attenuation of surface acoustic waves. Phys. Rev. Lett. {\bf 116}, 198001 (2016).

\bibitem{PichardPRB} H.  Pichard, A.  Duclos, J.-P.  Groby, and V.  Tournat,  Two-dimensional discrete granular phononic crystal for shear wave control. Phys. Rev. B {\bf 86}, 134307 (2012)

\bibitem{AlleinAPL} F.  Allein, V.  Tournat, V.~E.  Gusev, and G.  Theocharis, Tunable magneto-granular phononic crystals. Appl. Phys. Lett.  {\bf 108}, 161903 (2016).

\bibitem{TournatNJP} V. Tournat, I. P{\`e}rez-Arjona, A. Merkel, V Sanchez-Morcillo, and V. Gusev, Elastic waves in phononic monolayer granular membranes. New J. Phys. {\bf 13}, 073042 (2011). 

\bibitem{ZhengUltrasonics} L.-Y. Zheng, H. Pichard, V. Tournat, G. Theocharis, and V. Gusev, Zero-frequency and slow elastic modes in phononic monolayer granular membranes. Ultrasonics {\bf 69}, 201–214 (2016).

\bibitem{ZhengPRB} L.-Y. Zheng, G. Theocharis, V. Tournat, and V. Gusev, Quasitopological rotational waves in mechanical granular graphene. Phys. Rev. B {\bf 97}, 060101(R) (2018).

\bibitem{GillesPRL} B. Gilles, and C. Coste,  Low-frequency behavior of beads constrained on a Lattice. Phys. Rev. Lett. {\bf 90}, 174302 (2003).

\bibitem{CostePRE2008} C. Coste, and B. Gilles, Sound propagation in a constrained lattice of beads: High-frequency behavior and dispersion relation. Phys. Rev. E {\bf 77}, 021302 (2008).

\bibitem{PrlChiara} 
A. Leonard, C. Daraio,  Stress wave anisotropy in centered square highly nonlinear granular systems. Phys. Rev. Lett.  {\bf108}, 214301, (2012).

\bibitem{ZhuPRB} H. Zhu, T.-W. Liu, and F. Semperlotti, Design and experimental observation of valley-Hall edge states in diatomic-graphene-like elastic waveguides. Phys. Rev. B {\bf 97}, 174301 (2018).

\bibitem{Johnson} K. L. Johnson, Contact Mechanics. Cambridge Univ. Press, 1985.

\bibitem{Mindlin} R. D. Mindlin, Compliance of elastic bodies in contact. Journal of Applied Mechanics, {\bf 16}, 259 (1949).

\bibitem{BoechlerNat} N. Boechler, G. Theocharis, and C. Daraio, Bifurcation-based acoustic switching and rectification. Nature Materials {\bf 10}, 665–668 (2011).

%
%
%
%
%
\bibitem{Qian} 
K. Qian, D. J. Apigo, C. Prodan, Y. Barlas, and E. Prodan,  Topology of the valley-Chern effect. Phys. Rev. B {\bf 98}, 155138 (2018).

\bibitem{Babaee} 
S. Babaee, J. T. B. Overvelde, E. R. Chen, V. Tournat, and K. Bertoldi, Reconfigurable origami-inspired acoustic waveguides. Sci. Adv. {\bf 2},  e1601019 (2016).

\bibitem{Deng} 
B. Deng, P. Wang. Q. He, V. Tournat, and K. Bertoldi, Metamaterials with amplitude gaps for elastic solitons. Nat. Commun. {\bf 9}, 3410 (2018).

\bibitem{Chong} 
C. Chong, M. A Porter, P. G. Kevrekidis, and C Daraio, Nonlinear coherent structures in granular crystals. J. Phys.: Condens. Matter {\bf 29}, 413002 (2017).

\bibitem{Nesterenko}
V.F. Nesterenko, Dynamics of heterogeneous materials. Springer-Verlag, New York, 2001.

\bibitem{CostePRE} C. Coste, E. Falcon, and S. Fauve, Solitary waves in a chain of beads under Hertz contact. Phys. Rev. E {\bf 56}, 6104-6117 (1997).

\bibitem{JobPRL} S. Job, F. Melo, A. Sokolow, and S. Sen, How hertzian solitary waves interact with boundaries in a 1D granular medium. Phys. Rev. Lett. {\bf 94}, 178002 (2005).

\bibitem{LeonardExpMech}A. Leonard, F. Fraternali, and C. Daraio, Directional wave propagation in a highly nonlinear square packing of spheres. Experimental Mechanics  {\bf 53}, 327–337 (2013).

\bibitem{CabaretPRL} J.  Cabaret, P.  B{\'e}quin, G.  Theocharis, V.  Andreev, V.~E.  Gusev, and V.
  Tournat, Nonlinear hysteretic torsional waves. Phys. Rev. Lett. {\bf 115}, 054301 (2015).
  

  
  \bibitem{BoechlerPRL} N. Boechler, G. Theocharis, S. Job, P. G. Kevrekidis, Mason A. Porter, and C. Daraio, Discrete breathers in one-dimensional diatomic granular crystals. Phys. Rev. Lett. {\bf 104}, 244302 (2010).

\bibitem{TheocharisPRE} G. Theocharis, M. Kavousanakis, P. G. Kevrekidis, C. Daraio, M. A. Porter, and I. G. Kevrekidis, Localized breathing modes in granular crystals with defects. Phys. Rev. E {\bf 80}, 066601 (2009).

\end{thebibliography}
\end{document}